\documentclass[a4paper,ngerman,magyar,english]{article}
\usepackage{mathpazo}
\usepackage[T1]{fontenc}
\usepackage[latin2,latin9]{inputenc}
\usepackage{units}
\usepackage{amsthm}
\usepackage{amsmath}
\usepackage{amssymb}
\usepackage{graphicx}

\makeatletter

\pdfpageheight\paperheight
\pdfpagewidth\paperwidth

\newenvironment{lyxlist}[1]
{\begin{list}{}
{\settowidth{\labelwidth}{#1}
 \setlength{\leftmargin}{\labelwidth}
 \addtolength{\leftmargin}{\labelsep}
 }}
{\end{list}}
\theoremstyle{plain}
\newtheorem{thm}{\protect\theoremname}

\allowdisplaybreaks[0]

\newcounter{meg}

\newcommand{\R}{%
\refstepcounter{meg}%
{\vskip6pt \noindent \bf Remark \themeg.\ \ \ }%
}

\makeatother

\usepackage{babel}
  \addto\captionsenglish{\renewcommand{\theoremname}{Theorem}}
  \addto\captionsmagyar{\renewcommand{\theoremname}{Tétel}}
  \addto\captionsngerman{\renewcommand{\theoremname}{Theorem}}
\providecommand{\theoremname}{Theorem}

\begin{document}

\title{On the formal statement of the special principle of relativity}

\author{Márton Gömöri {\normalsize and} László E. Szabó\emph{\small }\\
\emph{\small Department of Logic, Institute of Philosophy}\\
\emph{\small Eötvös University, Budapest}\\
\emph{\small http://phil.elte.hu/logic}}

\date{~}
\maketitle
\begin{abstract}
The aim of the paper is to develop a proper mathematical formalism
which can help to clarify the necessary conceptual plugins to the
special principle of relativity and leads to a deeper understanding
of the principle in its widest generality.
\end{abstract}

\section{Introduction}

\noindent The aim of this paper is to spell out the (special) relativity
principle (RP) in a precise mathematical form. There are various verbal
formulations of the principle. In its shortest form it says that {}``All
the laws of physics take the same form in any inertial frame of reference.''
The laws of physics \emph{in }a reference frame\emph{ $K$} are meant
to be the laws of physics as they are ascertained by an observer being
at rest relative to the reference frame $K$; less anthropomorphically,
as they appear in the results of the measurements, such that both
the measuring equipments and the objects to be measured are co-moving
with $K$. 

For example, consider the following simple application of the principle
in Einstein's 1905 paper: 
\begin{quotation}
\noindent Let there be given a\emph{ stationary} rigid rod; and let
its length be $l$ as \emph{measured by a measuring-rod which is also
stationary}. We now imagine the axis of the rod lying along the axis
of $x$ of the stationary system of co-ordinates, and that a\emph{
}uniform\emph{ }motion of parallel translation with velocity $v$
along the axis of $x$ in the direction of increasing $x$ is then
imparted to the rod. We now inquire as to the length of the \emph{moving}
rod, and imagine its length to be ascertained by the following \emph{two}
operations:
\begin{lyxlist}{0.0.0}
\item [{(a)}] The observer \emph{moves together with the given measuring-rod
and the rod to be measured}, and \emph{measures the length} of the
rod directly by superposing the measuring-rod, \emph{in just the same
way} \emph{as if all three were at rest}.
\item [{(b)}] By means of \emph{stationary} clocks set up in the \emph{stationary}
system and synchronizing in accordance with {[}the light-signal synchronization{]},
the observer ascertains at what points of the \emph{stationary} system
the two ends of the rod to be measured are located at a definite time.
The distance between these two points, measured by the measuring-rod
already employed, which in this case is \emph{at rest}, is also a
length which may be designated {}``the length of the rod.'' 
\end{lyxlist}
\noindent \emph{In accordance with the principle of relativity} the
length \emph{to be discovered by the operation (a)}---we will call
it {}``the length of the rod in the moving system''---must be equal
to the length $l$ of \emph{the stationary} rod. 

The length to be discovered by the operation (b) we will call {}``the
length of the (\emph{moving}) rod in the \emph{stationary} system.''
This we shall determine \emph{on the basis of our two} \emph{principles},
and we shall find that it \emph{differs} from $l$. {[}all italics
added{]}
\end{quotation}
\inputencoding{latin2}%
\noindent \inputencoding{latin9}Thus, with the following formulation
(\inputencoding{latin2}\foreignlanguage{magyar}{Szabó}\inputencoding{latin9}~2004)
one can express in more detail how the principle is actually understood:
\begin{lyxlist}{00.00.0000}
\item [{(RP)}] The physical description of the behavior of a system co-moving
as a whole with an inertial frame\emph{ $K$}, expressed in terms
of the results of measurements obtainable by means of measuring equipments
co-moving with\emph{ $K$, }takes the same form as the description
of the similar behavior of the same system when it is co-moving with
another inertial frame\emph{ $K'$}, expressed in terms of the measurements
with the same equipments when they are co-moving with\emph{ $K'$}.
\end{lyxlist}
Our main concern in this paper is to unpack the verbal statement (RP)
and to provide its general mathematical formulation. We are hopeful
that the formalism we develop here helps to clarify the required conceptual
plugins to the RP and leads to a deeper understanding of the principle
in its widest generality.

\section{The statement of the RP \label{sec:The-statement-of}}

\noindent Consider an arbitrary collection of physical quantities
$\xi_{1},\xi_{2},\ldots\xi_{n}$ in $K$, operationally defined by
means of some\emph{ }operations with some\emph{ }equipments being
at rest in $K$. Let $\xi'_{1},\xi'_{2},\ldots\xi'_{n}$ denote another
collection of physical quantities that are defined by the \emph{same}
\emph{operations} with the \emph{same} \emph{equipments}, but \emph{in
different state of motion}, namely, in which they are all moving with
constant velocity $\mathbf{V}$ relative to $K$, co-moving with\emph{
$K'$}. Since, for all $i=1,2,\ldots n$, both $\xi_{i}$ and $\xi'_{i}$
are measured by the same equipment---although in different physical
conditions---with the same pointer scale, it is plausible to assume
that the possible values of $\xi_{i}$ and $\xi'_{i}$ range over
the same $\sigma_{i}\subseteq\mathbb{R}$. We introduce the following
notation: $\Sigma={\displaystyle \times_{i=1}^{n}\sigma_{i}}\subseteq\mathbb{R}^{n}$.

It must be emphasized that quantities $\xi_{1},\xi_{2},\ldots\xi_{n}$
and $\xi'_{1},\xi'_{2},\ldots\xi'_{n}$ are, a priori, \emph{different}
physical quantities, due to the fact that the operations by which
the quantities are defined are performed under \emph{different} physical
conditions; with measuring equipments of \emph{different} states of
motion. Any objective (non-conventional) relationship between them
must be a contingent law of nature. Thus, the same numeric values,
say, $(5,12,\ldots61)\in\mathbb{R}^{n}$ correspond to different states
of affairs when $\xi_{1}=5,\xi_{2}=12,\ldots\xi_{n}=61$ versus $\xi'_{1}=5,\xi'_{2}=12,\ldots\xi'_{n}=61$.
Consequently, $\left(\xi_{1},\xi_{2},\ldots\xi_{n}\right)$ and $\left(\xi'_{1},\xi'_{2},\ldots\xi'_{n}\right)$
are \emph{not} elements of the \emph{same }{}``space of physical
quantities''; although the numeric values of the physical quantities,
in both cases, can be represented in $\Sigma={\displaystyle \times_{i=1}^{n}\sigma_{i}}\subseteq\mathbb{R}^{n}$.

Mathematically, one can express this fact by means of two different
$n$-dimensional manifolds, $\Omega$ and $\Omega'$, each covered
by one global coordinate system, $\phi$ and $\phi'$ respectively,
such that $\phi:\Omega\rightarrow\Sigma$ assigns to every point of
$\Omega$ one of the possible $n$-tuples of numerical values of physical
quantities $\xi_{1},\xi_{2},\ldots\xi_{n}$ and $\phi':\Omega'\rightarrow\Sigma$
assigns to every point of $\Omega'$ one of the possible $n$-tuples
of numerical values of physical quantities $\xi'_{1},\xi'_{2},\ldots\xi'_{n}$
\begin{figure}
\begin{centering}
\includegraphics[width=0.7\columnwidth]{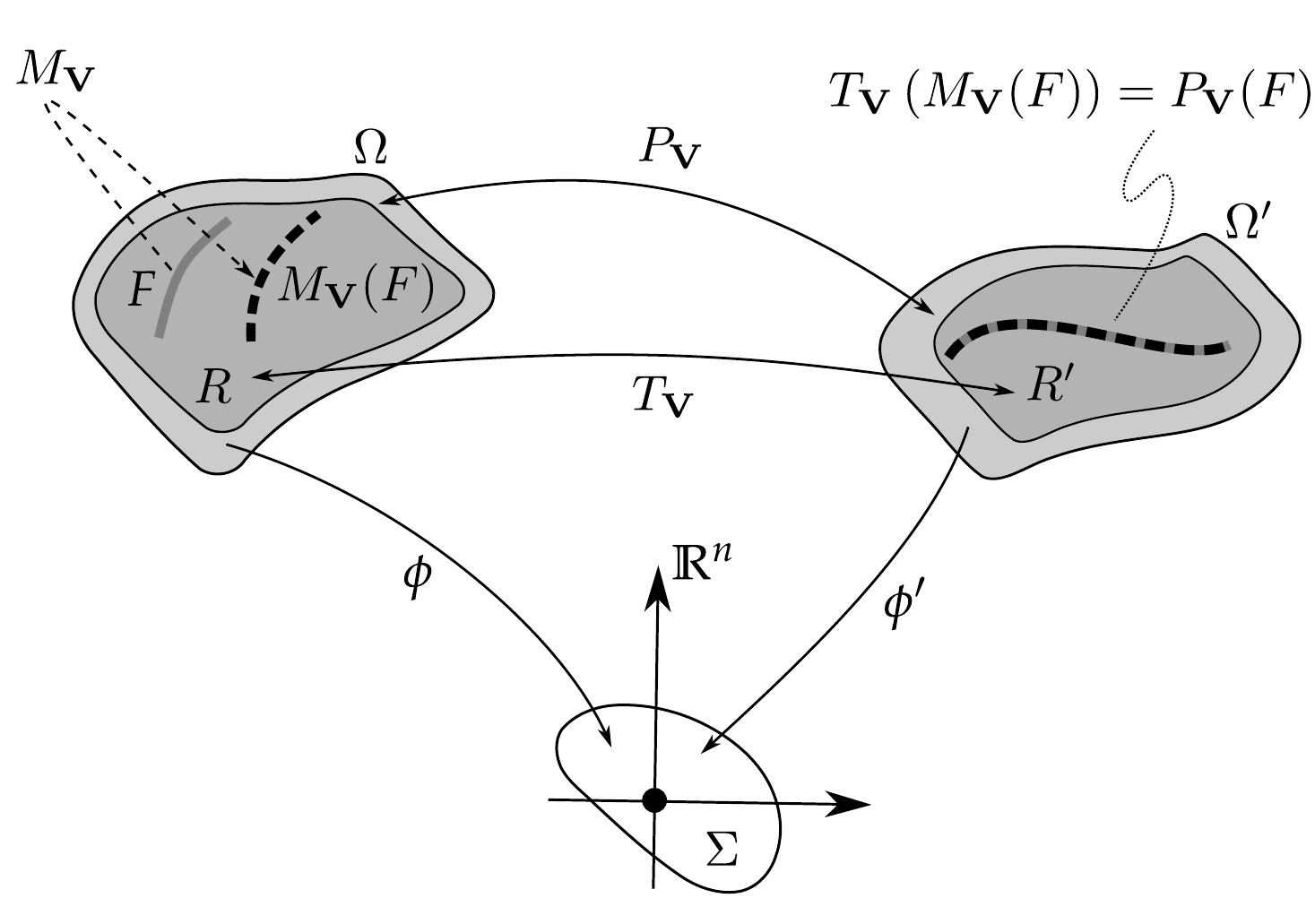}
\par\end{centering}

\caption{The relativity principle}
\label{Flo:relativity}
\end{figure}
 (Fig.~\ref{Flo:relativity}). In this way, a point $\omega\in\Omega$
represents the \emph{class} of physical constellations in which the
quantities $\xi_{1},\xi_{2},\ldots\xi_{n}$ take the values $\xi_{1}=\phi_{1}(\omega),\xi_{2}=\phi_{2}(\omega),\ldots\xi_{n}=\phi_{n}(\omega)$;
similarly, a point $\omega'\in\Omega'$ represents the physical constellation
characterized by $\xi'_{1}=\phi'_{1}(\omega'),\xi'_{2}=\phi'_{2}(\omega'),\ldots\xi'_{n}=\phi'_{n}(\omega')$.%
\footnote{$\phi_{i}=\pi_{i}\circ\phi$, where $\pi_{i}$ is the $i$-th coordinate
projection in $\mathbb{R}^{n}$.%
} Again, these physical constellations are generally different, even
in case of $\phi(\omega)=\phi'(\omega')\in\mathbb{R}^{n}$.

In the above sense, the points of $\Omega$ and the points of $\Omega'$
range over all possible value combinations of physical quantities
$\xi_{1},\xi_{2},\ldots\xi_{n}$ and $\xi'_{1},\xi'_{2},\ldots\xi'_{n}$.
It might be the case however that some combinations are impossible,
in the sense that they never come to existence in the physical world.
Let us denote by $R\subseteq\Omega$ and $R'\subseteq\Omega'$ the
physically admissible parts of $\Omega$ and $\Omega'$. Note that
$\phi(R)$ is not necessarily identical with $\phi'(R')$.%
\footnote{One can show however that $\phi(R)=\phi'(R')$ if the RP, that is
(\ref{eq:RP-math1}), holds.%
}

We shall use a bijection $P_{\mathbf{V}}:\Omega\rightarrow\Omega'$
({}``putting primes''; Bell 1987, p.~73) defined by means of the
two coordinate maps $\phi$ and $\phi'$: 
\begin{equation}
P_{\mathbf{V}}\overset{^{def}}{=}\left(\phi'\right)^{-1}\circ\phi\label{eq:F'}
\end{equation}

In contrast with $P_{\mathbf{V}}$, we now introduce the concept of
what we call the {}``transformation'' of physical quantities. It
is conceived as a bijection 
\begin{equation}
T_{\mathbf{V}}:\Omega\supseteq R\rightarrow R'\subseteq\Omega'
\end{equation}
determined by the contingent fact that whenever a physical constellation
belongs to the class represented by some $\omega\in R$ then it also
belongs to the class represented by $T_{\mathbf{V}}(\omega)\in R'$,
and vice versa. Since $\xi_{1},\xi_{2},\ldots\xi_{n}$ can be various
physical quantities in the various contexts, nothing guarantees that
such a bijection exists. We assume however the existence of $T_{\mathbf{V}}$.

\R \label{meg:peldak}It is worthwhile to consider several examples.
\begin{lyxlist}{00.00.0000}
\item [{(a)}] Let $\left(\xi_{1},\xi_{2}\right)$ be $\left(p,T\right)$,
the pressure and the temperature of a given (equilibrium) gas; and
let $\left(\xi'_{1},\xi'_{2}\right)$ be $\left(p',T'\right)$, the
pressure and the temperature of the same gas, measured by the moving
observer in $K'$. In this case, there exists a one-to-one $T_{\mathbf{V}}$:
\begin{eqnarray}
p' & = & p\label{eq:nyomas}\\
T' & = & T\gamma^{-1}\label{eq:homerseklet}
\end{eqnarray}
where $\gamma=\left(1-\frac{V^{2}}{c^{2}}\right)^{-\frac{1}{2}}$
(Tolman 1949, pp. 158--159).%
\footnote{There is a debate over the proper transformation rules (Georgieu 1969).%
}~A point $\omega\in\Omega$ of coordinates, say, $p=101325$ and
$T=300$ (in units $Pa$ and $^{\circ}K$) represents the class of
physical constellations---the class of possible worlds---in which
the gas in question has pressure of $101325\, Pa$ and temperature
of $300\,{}^{\circ}K$. Due to (\ref{eq:homerseklet}), this class
of physical constellations is different from the one represented by
$P_{\mathbf{V}}\left(\omega\right)\in\Omega'$ of coordinates $p'=101325$
and $T'=300$; but it is identical to the class of constellations
represented by $T_{\mathbf{V}}\left(\omega\right)\in\Omega'$ of coordinates
$p'=101325$ and $T'=300\gamma^{-1}$.
\item [{(b)}] Let $\left(\xi_{1},\xi_{2},\ldots\xi_{10}\right)$ be $\left(t,x,y,z,E_{x},E_{y},E_{z},r_{x},r_{y},r_{z}\right)$,
the time, the space coordinates where the electric field strength
is taken, the three components of the field strength, and the space
coordinates of a particle. And let $\left(\xi'_{1},\xi'_{2},\ldots\xi'_{10}\right)$
be $\left(t',x',y',z',E'_{x},E'_{y},E'_{z},r'_{x},r'_{y},r'_{z}\right)$,
the similar quantities obtainable by means of measuring equipments
co-moving with $K'$. In this case, there is no suitable one-to-one
$T_{\mathbf{V}}$, as the electric field strength in $K$ does not
determine the electric field strength in $K'$, and vice versa.
\item [{(c)}] Let $\left(\xi_{1},\xi_{2},\ldots\xi_{13}\right)$ be $\left(t,x,y,z,E_{x},E_{y},E_{z},B_{x},B_{y},B_{z},r_{x},r_{y},r_{z}\right)$
and let $\left(\xi'_{1},\xi'_{2},\ldots\xi'_{13}\right)$ be $\left(t',x',y',z',E'_{x},E'_{y},E'_{z},B'_{x},B'_{y},B'_{z},r'_{x},r'_{y},r'_{z}\right)$,
where $B_{x},B_{y},B_{z}$ and $B'_{x},B'_{y},B'_{z}$ are the magnetic
field strengths in $K$ and $K'$. In this case, in contrast with
(b), the well known Lorentz transformations of the spatio-temporal
coordinates and the electric and magnetic field strengths constitute
a proper one-to-one $T_{\mathbf{V}}$.\hfill{} $\lrcorner$\medskip{}

\end{lyxlist}
\noindent Next we turn to the general formulation of the concept of
the \emph{description of a particular behavior} of a physical system,
say, in $K$. We are probably not far from the truth if we assume
that such a description is, in its most abstract sense, a \emph{relation}
between physical quantities $\xi_{1},\xi_{2},\ldots\xi_{n}$; in other
words, it can be given as a subset $F\subset R$. 

\R Consider the above example~(a) in Remark~\ref{meg:peldak}.
An isochoric process of the gas can be described by the subset $F$
that is, in coordinates, determined by the following single equation:
\begin{equation}
F\,\,\,\left\{ p=\kappa T\right.\label{eq:kappa}
\end{equation}
with a certain constant $\kappa$. 

To give another example, consider the case (b). The relation $F$
given by equations 
\begin{equation}
F\,\,\,\left\{ \begin{alignedat}{1}E_{x} & =E_{0}\\
E_{y} & =0\\
E_{z} & =0\\
r_{x} & =x_{0}+v_{0}t\\
r_{y} & =0\\
r_{z} & =0
\end{alignedat}
\right.
\end{equation}
with some specific values of $E_{0},x_{0},v_{0}$ describes a neutral
particle moving with constant velocity in a static homogeneous electric
field. \hfill{} $\lrcorner$\medskip{}

\noindent Of course, one may not assume that an arbitrary relation
$F\subset R$ has physical meaning. Let $\mathcal{E}\subset2^{R}$
be the set of those $F\subset R$ which describe a particular behavior
of the system. We shall call $\mathcal{E}$ the\emph{ set of equations}
describing the physical system in question. The term is entirely justified.
In practical calculations, two systems of equations are regarded to
be equivalent if and only if they have the same solutions. Therefore,
a system of equations can be identified with the set of its solutions.
In general, the equations can be algebraic equations, ordinary and
partial integro-differential equations, linear and nonlinear, whatever.
So, in its most abstract sense, a system of equations is a set of
subsets of $R$. 

Now, consider the following subsets%
\footnote{We denote the map of type $\Omega\rightarrow\Omega'$ and its direct
image maps of type $2^{\Omega}\rightarrow2^{\Omega'}$ and $2^{2^{\Omega}}\rightarrow2^{2^{\Omega'}}$
or their restrictions by the same symbol. %
} of $\Omega'$, determined by an $F\in\mathcal{E}$:
\begin{lyxlist}{0000.0000.00}
\item [{$P_{\mathbf{V}}(F)\subseteq\Omega'$}] which formally is the {}``primed
$F$'', that is a relation of exactly the same {}``form'' as $F$,
but in the primed variables $\xi'_{1},\xi'_{2},\ldots\xi'_{n}$. Note
that relation $P_{\mathbf{V}}(F)$ does not necessarily describe a
true physical situation, as it can be not realized in nature.
\item [{$T_{\mathbf{V}}(F)\subseteq R'$}] which is the same description
of the same physical situation as $F$, but \emph{expressed} in the
primed variables.
\end{lyxlist}
We need one more concept. The RP is about the connection between two
situations: one is in which the system, as a whole, is at rest relative
to inertial frame $K$, the other is in which the system shows the
similar behavior, but being in a collective motion relative to $K$,
co-moving with $K'$. In other words, we assume the existence of a
map $M_{\mathbf{V}}:\,\mathcal{E}\rightarrow\mathcal{E}$, assigning
to each $F\in\mathcal{E}$, stipulated to describe the situation in
which the system is co-moving as a whole with inertial frame $K$,
another relation $M_{\mathbf{V}}(F)\in\mathcal{E}$, describing the
similar behavior of the same system when it is, as a whole, co-moving
with inertial frame\emph{ $K'$}, that is, when it is in a collective
motion with velocity $\mathbf{V}$ relative to $K$. 

Now, applying all these concepts, what the RP states is the following:
\begin{equation}
T_{\mathbf{V}}\left(M_{\mathbf{V}}(F)\right)=P_{\mathbf{V}}(F)\,\,\,\,\mbox{ for all }F\in\mathcal{E}\label{eq:RP-math1}
\end{equation}
or equivalently, 
\begin{equation}
P_{\mathbf{V}}(F)\subset R'\mbox{ and }M_{\mathbf{V}}(F)=T_{\mathbf{V}}^{-1}\left(P_{\mathbf{V}}(F)\right)\,\,\,\,\mbox{ for all }F\in\mathcal{E}\label{eq:RP-math1a}
\end{equation}

\R\label{Meg:kontingencia}Notice that, for a given fixed $F$, everything
on the right hand side of the equation in (\ref{eq:RP-math1a}), $P_{\mathbf{V}}$
and $T_{\mathbf{V}}$, are determined \emph{only} \emph{by} the physical
behaviors of \emph{the} \emph{measuring equipments} when they are
in various states of motion. In contrast, the meaning of the left
hand side, $M_{\mathbf{V}}(F)$, depends on the physical behavior
of \emph{the object physical system} described by $F$ and $M_{\mathbf{V}}(F)$,
when it is in various states of motion. That is to say, the two sides
of the equation reflect the behaviors\emph{ }of\emph{ different parts}
of the physical reality; and the RP expresses a law-like regularity
between the behaviors\emph{ }of\emph{ }these different parts\emph{.}\hfill{}
$\lrcorner$

\R\label{meg:example}Let us illustrate these concepts with a well-known
textbook example of a static versus uniformly moving charged particle.
The static field of a charge $q$ being at \emph{rest} at point $(x_{0},y_{0},z_{0})$
in $K$ is the following:
\begin{equation}
F\,\,\,\left\{ \begin{alignedat}{1}E_{x} & =\frac{q\left(x-x_{0}\right)}{\left(\left(x-x_{0}\right)^{2}+\left(y-y_{0}\right)^{2}+\left(z-z_{0}\right)^{2}\right)^{\nicefrac{3}{2}}}\\
E_{y} & =\frac{q\left(y-y_{0}\right)}{\left(\left(x-x_{0}\right)^{2}+\left(y-y_{0}\right)^{2}+\left(z-z_{0}\right)^{2}\right)^{\nicefrac{3}{2}}}\\
E_{z} & =\frac{q\left(z-z_{0}\right)}{\left(\left(x-x_{0}\right)^{2}+\left(y-y_{0}\right)^{2}+\left(z-z_{0}\right)^{2}\right)^{\nicefrac{3}{2}}}\\
B_{x} & =0\\
B_{y} & =0\\
B_{z} & =0
\end{alignedat}
\right.\label{eq:Coulomb field}
\end{equation}

The stationary field of a charge $q$ \emph{moving} at constant velocity
$\mathbf{V}=\left(V,0,0\right)$ relative to $K$ can be obtained
by solving the equations of electrodynamics (in $K$) with the time-depending
source (for example, Jackson 1999, pp. 661--665): 

\begin{equation}
M_{\mathbf{V}}(F)\,\,\,\left\{ \begin{alignedat}{1}E_{x} & =\frac{qX_{0}}{\left(X_{0}^{2}+\left(y-y_{0}\right)^{2}+\left(z-z_{0}\right)^{2}\right)^{\nicefrac{3}{2}}}\\
E_{y} & =\frac{\gamma q\left(y-y_{0}\right)}{\left(X_{0}^{2}+\left(y-y_{0}\right)^{2}+\left(z-z_{0}\right)^{2}\right)^{\nicefrac{3}{2}}}\\
E_{z} & =\frac{\gamma q\left(z-z_{0}\right)}{\left(X_{0}^{2}+\left(y-y_{0}\right)^{2}+\left(z-z_{0}\right)^{2}\right)^{\nicefrac{3}{2}}}\\
B_{x} & =0\\
B_{y} & =-c^{-2}VE_{z}\\
B_{z} & =c^{-2}VE_{y}
\end{alignedat}
\right.\label{eq:Coulomb-mozgo}
\end{equation}
where where $(x_{0},y_{0},z_{0})$ is the initial position of the
particle at $t=0$, $X_{0}=\gamma\left(x-\left(x_{0}+Vt\right)\right)$.

Now, we form the same expressions as (\ref{eq:Coulomb field}) but
in the \emph{primed} variables of the co-moving reference frame $K'$:
\begin{equation}
P_{\mathbf{V}}\left(F\right)\,\,\,\left\{ \begin{alignedat}{1}E'_{x} & =\frac{q'\left(x'-x'_{0}\right)}{\left(\left(x'-x'_{0}\right)^{2}+\left(y'-y'_{0}\right)^{2}+\left(z'-z'_{0}\right)^{2}\right)^{\nicefrac{3}{2}}}\\
E'_{y} & =\frac{q'\left(y'-y'_{0}\right)}{\left(\left(x'-x'_{0}\right)^{2}+\left(y'-y'_{0}\right)^{2}+\left(z'-z'_{0}\right)^{2}\right)^{\nicefrac{3}{2}}}\\
E'_{z} & =\frac{q'\left(z'-z'_{0}\right)}{\left(\left(x'-x'_{0}\right)^{2}+\left(y'-y'_{0}\right)^{2}+\left(z'-z'_{0}\right)^{2}\right)^{\nicefrac{3}{2}}}\\
B'_{x} & =0\\
B'_{y} & =0\\
B'_{z} & =0
\end{alignedat}
\right.\label{eq:Coulomb-vesszos}
\end{equation}
By means of the Lorentz transformation rules of the space-time coordinates,
the field strengths and the electric charge (e.g. Tolman 1949), one
can express (\ref{eq:Coulomb-vesszos}) in terms of the original variables
of $K$:

\begin{equation}
T_{\mathbf{V}}^{-1}\left(P_{\mathbf{V}}(F)\right)\,\,\,\left\{ \begin{alignedat}{1}E_{x} & =\frac{qX_{0}}{\left(X_{0}^{2}+\left(y-y_{0}\right)^{2}+\left(z-z_{0}\right)^{2}\right)^{\nicefrac{3}{2}}}\\
E_{y} & =\frac{\gamma q\left(y-y_{0}\right)}{\left(X_{0}^{2}+\left(y-y_{0}\right)^{2}+\left(z-z_{0}\right)^{2}\right)^{\nicefrac{3}{2}}}\\
E_{z} & =\frac{\gamma q\left(z-z_{0}\right)}{\left(X_{0}^{2}+\left(y-y_{0}\right)^{2}+\left(z-z_{0}\right)^{2}\right)^{\nicefrac{3}{2}}}\\
B_{x} & =0\\
B_{y} & =-c^{-2}VE_{z}\\
B_{z} & =c^{-2}VE_{y}
\end{alignedat}
\right.
\end{equation}
We find that the result is indeed the same as (\ref{eq:Coulomb-mozgo})
describing the field of the moving charge: $M_{\mathbf{V}}(F)=T_{\mathbf{V}}^{-1}\left(P_{\mathbf{V}}(F)\right)$.
That is to say, the RP seems to be true in this particular case. 

Reversely, \emph{assuming} that the particle + electromagnetic field
system satisfies the RP, that is, (\ref{eq:RP-math1a}) holds for
the equations of electrodynamics, one can \emph{derive} the stationary
field of a uniformly moving point charge (\ref{eq:Coulomb-mozgo})
from the static field (\ref{eq:Coulomb field}).\hfill{} $\lrcorner$

\section{Covariance}

\noindent Now we have a strict mathematical formulation of the RP
for a physical system described by a system of equations $\mathcal{E}$.
Remarkably, however, we still have not encountered the concept of
{}``covariance'' of equations $\mathcal{E}$. The reason is that
the RP and the covariance of equations $\mathcal{E}$ are not equivalent---in
contrast to what many believe. In fact, the logical relationship between
the two conditions is much more complex. To see this relationship
in more detail, we previously need to clarify a few things. 

Consider the following two sets: $P_{\mathbf{V}}(\mathcal{E})=\{P_{\mathbf{V}}(F)|F\in\mathcal{E}\}$
and $T_{\mathbf{V}}(\mathcal{E})=\{T_{\mathbf{V}}(F)|F\in\mathcal{E}\}$.
Since a system of equations can be identified with its set of solutions,
$P_{\mathbf{V}}(\mathcal{E})\subset2^{\Omega'}$ and $T_{\mathbf{V}}(\mathcal{E})\subset2^{R'}$
can be regarded as two systems of equations for functional relations
between $\xi'_{1},\xi'_{2},\ldots\xi'_{n}$. In the primed variables,
$P_{\mathbf{V}}(\mathcal{E})$ has {}``the same form'' as $\mathcal{E}$.
Nevertheless, it can be the case that $P_{\mathbf{V}}(\mathcal{E})$
does not express a true physical law, in the sense that its solutions
do not necessarily describe true physical situations. In contrast,
$T_{\mathbf{V}}(\mathcal{E})$ is nothing but $\mathcal{E}$ expressed
in variables\emph{ $\xi'_{1},\xi'_{2},\ldots\xi'_{n}$}. 

Now, covariance intuitively means that equations $\mathcal{E}$ {}``preserve
their forms against the transformation $T_{\mathbf{V}}$''. That
is, in terms of the formalism we developed: 
\begin{equation}
T_{\mathbf{V}}(\mathcal{E})=P_{\mathbf{V}}(\mathcal{E})\label{eq:kovi}
\end{equation}
or, equivalently,
\begin{equation}
P_{\mathbf{V}}(\mathcal{E})\subset2^{R'}\,\mbox{ and }\,\mathcal{E}=T_{\mathbf{V}}^{-1}\left(P_{\mathbf{V}}(\mathcal{E})\right)\label{eq:kovi-a}
\end{equation}

The first thing we have to make clear is that---even if we know or
presume that it holds---covariance (\ref{eq:kovi-a}) is obviously
\emph{not sufficient} for the RP (\ref{eq:RP-math1a}). For, (\ref{eq:kovi-a})
only guarantees the invariance of the set of solutions, $\mathcal{E}$,
against $T_{\mathbf{V}}^{-1}\circ P_{\mathbf{V}}$ , but it says nothing
about which solution of $\mathcal{E}$ corresponds to which solution.
In Bell's words:
\begin{quote}
Lorentz invariance alone shows that for any state of a system at rest
there is a corresponding `primed' state of that system in motion.
But it does not tell us that if the system is set anyhow in motion,
it will actually go into the 'primed' of the original state, rather
than into the `prime' of some \emph{other} state of the original system.
(Bell 1987, p.~75)
\end{quote}
While it is the very essence of the RP that the solution $M_{\mathbf{V}}(F)$,
describing the system \emph{in motion} relative to $K$, corresponds
to solution $T_{\mathbf{V}}^{-1}\circ P_{\mathbf{V}}(F)$. For example,
what we use in the above mentioned textbook derivation of the stationary
electromagnetic field of a uniformly moving point charge (end of Remark~\ref{meg:example})
is not the covariance of the equations---that would be not enough---but
statement~(\ref{eq:RP-math1a}), that is, what the RP claims about
the solutions of the equations in detail.

In a precise sense, covariance is not only not sufficient for the
RP, but it is \emph{not even necessary} 
\begin{figure}
\begin{centering}
\includegraphics[width=0.8\textwidth]{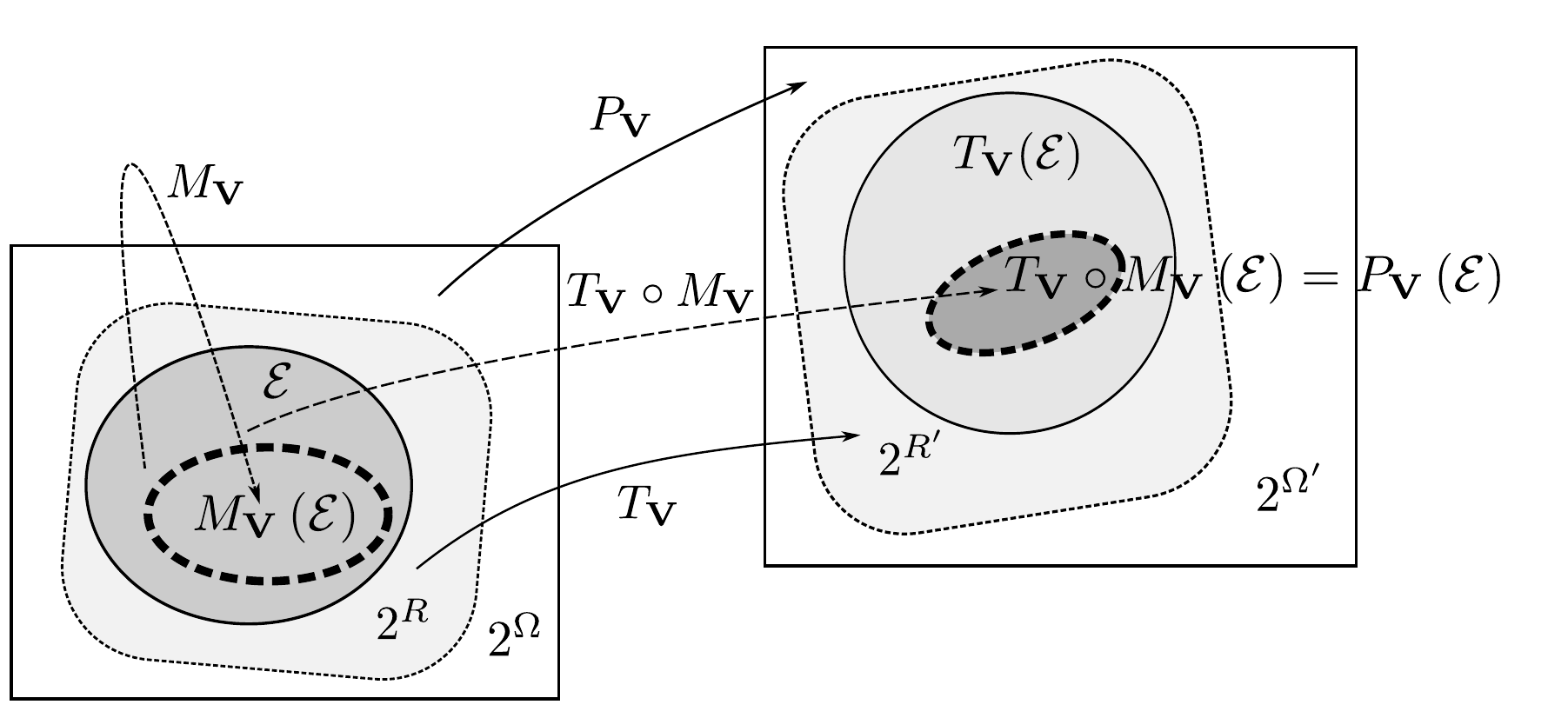}
\par\end{centering}

\caption{The RP only implies that $T_{\mathbf{V}}(\mathcal{E})\supseteq T_{\mathbf{V}}\circ M_{\mathbf{V}}\left(\mathcal{E}\right)=P_{\mathbf{V}}\left(\mathcal{E}\right)$.
Covariance of $\mathcal{E}$ would require that $T_{\mathbf{V}}(\mathcal{E})=P_{\mathbf{V}}(\mathcal{E})$,
which is generally not the case\label{fig:Relativity-principle-only}}
\end{figure}
 (Fig.~\ref{fig:Relativity-principle-only}). The RP only implies
that 
\begin{equation}
T_{\mathbf{V}}(\mathcal{E})\supseteq T_{\mathbf{V}}\left(M_{\mathbf{V}}\left(\mathcal{E}\right)\right)=P_{\mathbf{V}}\left(\mathcal{E}\right)
\end{equation}
 (\ref{eq:RP-math1}) implies (\ref{eq:kovi}) only if we have the
following \emph{extra} condition: 
\begin{equation}
M_{\mathbf{V}}\left(\mathcal{E}\right)=\mathcal{E}\label{eq:RE-kov2}
\end{equation}

\section{Initial and boundary conditions}

Let us finally consider the situation when the solutions of a system
of equations $\mathcal{E}$ are specified by some extra conditions---initial
and/or boundary value conditions, for example. In our general formalism,
an extra condition for $\mathcal{E}$ is a system of equations $\psi\subset2^{\Omega}$
such that there exists exactly one solution $\left[\psi\right]_{\mathcal{E}}$
satisfying both $\mathcal{E}$ and $\psi$. That is, $\mathcal{E}\cap\psi=\left\{ \left[\psi\right]_{\mathcal{E}}\right\} $,
where $\left\{ \left[\psi\right]_{\mathcal{E}}\right\} $ is a singleton
set. Since $\mathcal{E}\subset2^{R}$, without loss of generality
we may assume that $\psi\subset2^{R}$. 

Since $P_{\mathbf{V}}$ and $T_{\mathbf{V}}$ are injective, $P_{\mathbf{V}}\left(\psi\right)$
and $T_{\mathbf{V}}\left(\psi\right)$ are extra conditions for equations
$P_{\mathbf{V}}\left(\mathcal{E}\right)$ and $T_{\mathbf{V}}\left(\mathcal{E}\right)$
respectively, and we have 
\begin{eqnarray}
P_{\mathbf{V}}\left(\left[\psi\right]_{\mathcal{E}}\right) & = & \left[P_{\mathbf{V}}\left(\psi\right)\right]_{P_{\mathbf{V}}\left(\mathcal{E}\right)}\label{eq:atalakitas2}\\
T_{\mathbf{V}}\left(\left[\psi\right]_{\mathcal{E}}\right) & = & \left[T_{\mathbf{V}}\left(\psi\right)\right]_{T_{\mathbf{V}}\left(\mathcal{E}\right)}\label{eq:atalakitas}
\end{eqnarray}
for all extra conditions $\psi$ for $\mathcal{E}$. Similarly, if
$P_{\mathbf{V}}(\mathcal{E}),P_{\mathbf{V}}\left(\psi\right)\subset2^{R'}$
then $T_{\mathbf{V}}^{-1}\left(P_{\mathbf{V}}\left(\psi\right)\right)$
is an extra condition for $T_{\mathbf{V}}^{-1}\left(P_{\mathbf{V}}\left(\mathcal{E}\right)\right)$,
and 
\begin{equation}
\left[T_{\mathbf{V}}^{-1}\left(P_{\mathbf{V}}\left(\psi\right)\right)\right]_{T_{\mathbf{V}}^{-1}\left(P_{\mathbf{V}}\left(\mathcal{E}\right)\right)}=T_{\mathbf{V}}^{-1}\left(\left[P_{\mathbf{V}}(\psi)\right]_{P_{\mathbf{V}}(\mathcal{E})}\right)\label{eq:atalakitas4}
\end{equation}

Consider now a set of extra conditions $\mathcal{C}\subset2^{2^{R}}$.
Assume that $\mathcal{C}$ is a \emph{parametrizing set of extra conditions}
for $\mathcal{E}$; by which we mean that for all $F\in\mathcal{E}$
there exists exactly one $\psi\in\mathcal{C}$ such that $F=\left[\psi\right]_{\mathcal{E}}$;
in other words, 
\begin{equation}
\mathcal{C}\ni\psi\mapsto\left[\psi\right]_{\mathcal{E}}\in\mathcal{E}
\end{equation}
is a bijection. 

$M_{\mathbf{V}}:\mathcal{E}\rightarrow\mathcal{E}$ was introduced
as a map between solutions of $\mathcal{E}$. Now, as there is a one-to-one
correspondence between the elements of $\mathcal{C}$ and $\mathcal{E}$,
it generates a map $M_{\mathbf{V}}:\mathcal{C}\rightarrow\mathcal{C}$,
such that 
\begin{equation}
\left[M_{\mathbf{V}}(\psi)\right]_{\mathcal{E}}=M_{\mathbf{V}}\left(\left[\psi\right]_{\mathcal{E}}\right)\label{eq:atalakitas3}
\end{equation}

Thus, from \eqref{eq:atalakitas2} and \eqref{eq:atalakitas3}, the
RP, that is (\ref{eq:RP-math1}), has the following form:
\begin{equation}
T_{\mathbf{V}}\left(\left[M_{\mathbf{V}}(\psi)\right]_{\mathcal{E}}\right)=\left[P_{\mathbf{V}}(\psi)\right]_{P_{\mathbf{V}}(\mathcal{E})}\,\,\,\,\,\,\,\mbox{for all}\,\psi\in\mathcal{C}\label{eq:Lequivalent2}
\end{equation}
 or, equivalently, \eqref{eq:RP-math1a} reads
\begin{equation}
\left[P_{\mathbf{V}}(\psi)\right]_{P_{\mathbf{V}}(\mathcal{E})}\subset R'\,\mbox{ and }\,\left[M_{\mathbf{V}}(\psi)\right]_{\mathcal{E}}=T_{\mathbf{V}}^{-1}\left(\left[P_{\mathbf{V}}(\psi)\right]_{P_{\mathbf{V}}(\mathcal{E})}\right)\label{eq:Lequivalent2a}
\end{equation}

One might make use of the following theorem:
\begin{thm}
\label{thm:RV-ekvivalens} Assume that the system of equations $\mathcal{E}\subset2^{R}$
is covariant, that is, (\ref{eq:kovi}) is satisfied. Then, 
\begin{itemize}
\item [(i)] for all $\psi\in\mathcal{C}$, $T_{\mathbf{V}}\left(M_{\mathbf{V}}\left(\psi\right)\right)$
is an extra condition for the system of equations $P_{\mathbf{V}}\left(\mathcal{E}\right)$,
and, (\ref{eq:Lequivalent2}) is equivalent to the following condition:
\begin{equation}
\left[T_{\mathbf{V}}\left(M_{\mathbf{V}}(\psi)\right)\right]_{P_{\mathbf{V}}(\mathcal{E})}=\left[P_{\mathbf{V}}(\psi)\right]_{P_{\mathbf{V}}(\mathcal{E})}\label{eq:masodik-feltetel_a}
\end{equation}

\item [(ii)] for all $\psi\in\mathcal{C},P_{\mathbf{V}}\left(\psi\right)\subset2^{R'}$,
$T_{\mathbf{V}}^{-1}\left(P_{\mathbf{V}}\left(\psi\right)\right)$
is an extra condition for the system of equations $\mathcal{E}$ and
(\ref{eq:Lequivalent2a}) is equivalent to the following condition:
\begin{equation}
\left[M_{\mathbf{V}}(\psi)\right]_{\mathcal{E}}=\left[T_{\mathbf{V}}^{-1}\left(P_{\mathbf{V}}\left(\psi\right)\right)\right]_{\mathcal{E}}\label{eq:masodik-feltetel}
\end{equation}

\end{itemize}
\end{thm}
\begin{proof}
(i)~~~Obviously, $T_{\mathbf{V}}\left(\mathcal{E}\right)\cap T_{\mathbf{V}}\left(M_{\mathbf{V}}\left(\psi\right)\right)$
exists and is a singleton; and, due to \eqref{eq:kovi}, it is equal
to $P_{\mathbf{V}}\left(\mathcal{E}\right)\cap T_{\mathbf{V}}\left(M_{\mathbf{V}}\left(\psi\right)\right)$;
therefore this latter is a singleton, too. Applying \eqref{eq:atalakitas}
and \eqref{eq:kovi}, we have 
\begin{equation}
T_{\mathbf{V}}\left(\left[M_{\mathbf{V}}(\psi)\right]_{\mathcal{E}}\right)=\left[T_{\mathbf{V}}\left(M_{\mathbf{V}}\left(\psi\right)\right)\right]_{T_{\mathbf{V}}\left(\mathcal{E}\right)}=\left[T_{\mathbf{V}}\left(M_{\mathbf{V}}\left(\psi\right)\right)\right]_{P_{\mathbf{V}}\left(\mathcal{E}\right)}
\end{equation}
therefore, \eqref{eq:masodik-feltetel_a} implies \eqref{eq:Lequivalent2a}.

(ii)~~~Similarly, due to $P_{\mathbf{V}}\left(\psi\right)\subset2^{R'}$
and \eqref{eq:kovi-a}, $\mathcal{E}\cap T_{\mathbf{V}}^{-1}\left(P_{\mathbf{V}}\left(\psi\right)\right)$
exists and is a singleton. Applying \eqref{eq:atalakitas4} and \eqref{eq:kovi-a},
we have
\begin{equation}
T_{\mathbf{V}}^{-1}\left(\left[P_{\mathbf{V}}(\psi)\right]_{P_{\mathbf{V}}(\mathcal{E})}\right)=\left[T_{\mathbf{V}}^{-1}\left(P_{\mathbf{V}}\left(\psi\right)\right)\right]_{T_{\mathbf{V}}^{-1}\left(P_{\mathbf{V}}\left(\mathcal{E}\right)\right)}=\left[T_{\mathbf{V}}^{-1}\left(P_{\mathbf{V}}\left(\psi\right)\right)\right]_{\mathcal{E}}
\end{equation}
that is, \eqref{eq:masodik-feltetel} implies \eqref{eq:Lequivalent2a}.
\end{proof}
\R\label{meg:kovariancia-uj}Let us note a few important facts which
can easily be seen in the formalism we developed:
\begin{lyxlist}{00.00.0000}
\item [{(a)}] The covariance of a set of equations $\mathcal{E}$ does
\emph{not} imply the covariance of a subset of equations separately.
It is because a smaller set of equations corresponds to an $\mathcal{E}^{*}\subset2^{R}$
such that $\mathcal{E}\subset\mathcal{E}^{*}$; and it does not follow
from (\ref{eq:kovi}) that $T_{\mathbf{V}}(\mathcal{E}^{*})=P_{\mathbf{V}}(\mathcal{E}^{*})$.
\item [{(b)}] Similarly, the covariance of a set of equations $\mathcal{E}$
does \emph{not} guarantee the covariance of an arbitrary set of equations
which is only satisfactory to $\mathcal{E}$; for example, when the
solutions of $\mathcal{E}$ are restricted by some extra conditions.
Because from (\ref{eq:kovi}) it does not follow that $T_{\mathbf{V}}(\mathcal{E}^{*})=P_{\mathbf{V}}(\mathcal{E}^{*})$
for an arbitrary $\mathcal{E}^{*}\subset\mathcal{E}$.
\item [{(c)}] The same holds, of course, for the combination of cases (a)
and (b); for example, when we have a smaller set of equations $\mathcal{E}^{*}\supset\mathcal{E}$
together with some extra conditions $\psi\subset2^{R}$. For, (\ref{eq:kovi})
does not imply that $T_{\mathbf{V}}(\mathcal{E}^{*}\cap\psi)=P_{\mathbf{V}}(\mathcal{E}^{*}\cap\psi)$.
\item [{(d)}] However, covariance is guaranteed if a covariant set of equations
is restricted with a \emph{covariant} set of extra conditions; because
$T_{\mathbf{V}}(\mathcal{E})=P_{\mathbf{V}}(\mathcal{E})$ and $T_{\mathbf{V}}(\psi)=P_{\mathbf{V}}(\psi)$
trivially imply that $T_{\mathbf{V}}(\mathcal{E}\cap\psi)=P_{\mathbf{V}}(\mathcal{E}\cap\psi)$.\hfill{}
$\lrcorner$\medskip{}

\end{lyxlist}

\section{Concluding discussions and open problems\label{sec:Concluding-discussions-and}}

As we have seen, the notion of $M_{\mathbf{V}}$ plays a crucial role.
Formally, one could say, the RP is \emph{relative} to the definition
of $M_{\mathbf{V}}$; the physical content of the RP depends on how
this concept is physically understood. But, what does it mean to say
that a physical system is the same and of the same behavior as the
one described by $F$, except that it is, as a whole, in a collective
motion with velocity $\mathbf{V}$ relative to $K$? Without answering
this crucial question the RP is meaningless. 

In fact, the same question can be asked with respect to the definitions
of quantities $\xi'_{1},\xi'_{2},\ldots\xi'_{n}$---and, therefore,
with respect to the meanings of $T_{\mathbf{V}}$ and $P_{\mathbf{V}}$.
For, $\xi'_{1},\xi'_{2},\ldots\xi'_{n}$ are not simply arbitrary
variables assigned to reference frame $K'$, in one-to-one relations
with $\xi_{1},\xi_{2},\ldots\xi_{n}$, but the physical quantities
obtainable by means of the same operations with the same measuring
equipments as in the operational definitions of $\xi_{1},\xi_{2},\ldots\xi_{n}$,
except that everything is in a collective motion with velocity $\mathbf{V}$.
Therefore, we should know what we mean by {}``the same measuring
equipment but in collective motion''. From this point of view, it
does not matter whether the system in question is the object to be
observed or a measuring equipment involved in the observation. 

These questions can be answered only within the given physical context;
and, one must admit, in some situations the answers are non trivial
and ambiguous (cf. Szabó~2004). At this level of generality we only
want to point out two things.

First, whatever is the definition of $M_{\mathbf{V}}:\,\mathcal{E}\rightarrow\mathcal{E}$
in the given context, the following is a minimal requirement for it
to have the assumed physical meaning:
\begin{lyxlist}{00.00.0000}
\item [{(M)}] Relations $F\in\mathcal{E}$ must describe situations which\emph{
}can be meaningfully characterized as such in which the system as
a whole is at rest or in motion with some velocity relative to a frame
of reference. 
\end{lyxlist}
For example, in Remark~\textbf{\ref{meg:example}}, solutions (\ref{eq:Coulomb field})
and (\ref{eq:Coulomb-mozgo}) satisfy this condition, as in both cases
the system of the charged particle + electromagnetic field qualifies
as a system in collective rest or motion. The electromagnetic field
is in collective motion with the point charge of velocity $\mathbf{V}$
(Fig.~\ref{fig:The-stationary-field-1}) in the following sense:
\begin{figure}
\begin{centering}
\includegraphics[width=0.6\columnwidth]{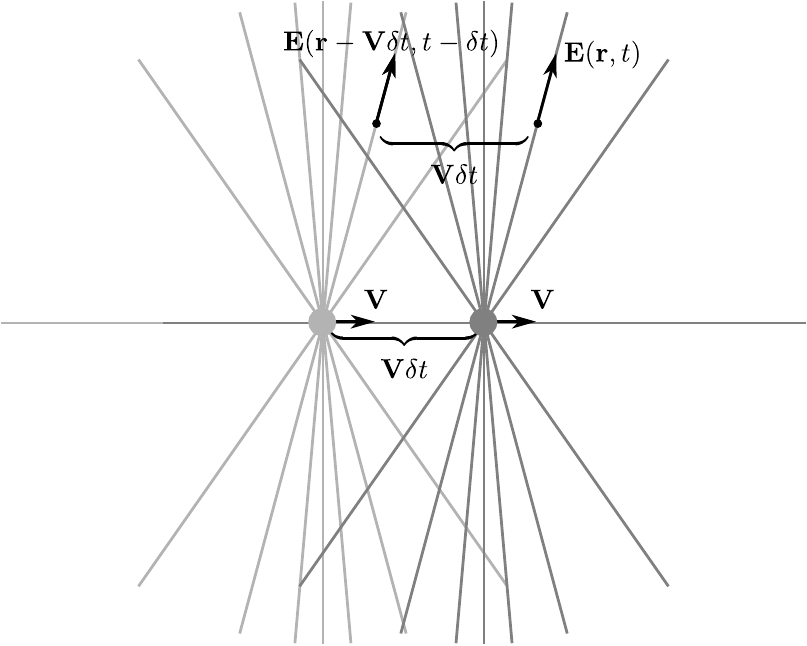}
\par\end{centering}

\caption{The stationary field of a uniformly moving point charge is in collective
motion together with the point charge \label{fig:The-stationary-field-1}}
\end{figure}
\begin{eqnarray}
\mathbf{E}(\mathbf{r},t) & = & \mathbf{E}(\mathbf{r}-\mathbf{V}\delta t,t-\delta t)\label{eq:mozgomezo-0-1}\\
\mathbf{B}(\mathbf{r},t) & = & \mathbf{B}(\mathbf{r}-\mathbf{V}\delta t,t-\delta t)\label{eq:mozgomezo-0-2}
\end{eqnarray}

Notice that requirement (M) says nothing about whether and how the
fact that the system as a whole is at rest or in motion with some
velocity is reflected in the solutions $F\in\mathcal{E}$. It does
not even require that this fact can be expressed in terms of $\xi_{1},\xi_{2},\ldots\xi_{n}$.
It only requires that each $F\in\mathcal{E}$ belong to a physical
situation in which it is meaningful to say---perhaps in terms of quantities
different from $\xi_{1},\xi_{2},\ldots\xi_{n}$---that the system
is at rest or in motion relative to a reference frame. How a concrete
physical situation can be characterized as such in which the system
is at rest or in motion is a separate problem, which can be discussed
in the particular contexts.

The second thing to be said about $M_{\mathbf{V}}(F)$ is that it
is a notion determined by the concrete physical context; but it is
\emph{not} equal to the {}``Lorentz boosted solution'' $T_{\mathbf{V}}^{-1}\left(P_{\mathbf{V}}(F)\right)$
\emph{by definition} ---as a little reflection shows:
\begin{lyxlist}{00.00.0000}
\item [{(a)}] In this case, (\ref{eq:RP-math1a}) would read 
\begin{equation}
T_{\mathbf{V}}^{-1}\left(P_{\mathbf{V}}(F)\right)=T_{\mathbf{V}}^{-1}\left(P_{\mathbf{V}}(F)\right)\label{eq:RP-tautologi}
\end{equation}
That is, the RP would become a tautology; a statement which is always
true, independently of any contingent fact of nature; independently
of the actual behavior of moving physical objects; and independently
of the actual empirical meanings of physical quantities $\xi'_{1},\xi'_{2},\ldots\xi'_{n}$.
But, the RP is supposed to be a fundamental \emph{law of nature}.
Note that a tautology is entirely different from a fundamental principle,
even if the principle is used as a fundamental hypothesis or fundamental
premise of a theory, from which one derives further physical statements.
For, a fundamental premise, as expressing a contingent fact of nature,
is potentially falsifiable by testing its consequences; a tautology
is not.
\item [{(b)}] Even if accepted, $M_{\mathbf{V}}(F)\overset{^{def}}{=}T_{\mathbf{V}}^{-1}\left(P_{\mathbf{V}}(F)\right)$
can provide physical meaning to $M_{\mathbf{V}}(F)$ only if we know
the meanings of $T_{\mathbf{V}}$ and $P_{\mathbf{V}}$, that is,
if we know the empirical meanings of the quantities denoted by $\xi'_{1},\xi'_{2},\ldots\xi'_{n}$.
But, the physical meaning of $\xi'_{1},\xi'_{2},\ldots\xi'_{n}$ are
obtained from the operational definitions: they are the quantities
obtained by {}``the same measurements with the same equipments when
they are, as a whole, co-moving with\emph{ $K'$ }with velocity $\mathbf{V}$
relative to $K$''. Symbolically, we need, priory, the concepts of
$M_{\mathbf{V}}(\xi_{i}\mbox{-}equipment\, at\, rest)$. And this
is a conceptual circularity: in order to have the concept of what
it is to be an $M_{\mathbf{V}}(brick\, at\, rest)$ the (size)' of
which we would like to ascertain, we need to have the concept of what
it is to be an $M_{\mathbf{V}}(measuring\, rod\, at\, rest)$---which
is exactly the same conceptual problem. 
\item [{(c)}] One might claim that we do not need to specify the concepts
of $M_{\mathbf{V}}(\xi_{i}\mbox{-}equipment\, at\, rest)$ in order
to know the \emph{values} of quantities $\xi'_{1},\xi'_{2},\ldots\xi'_{n}$
we obtain by the measurements with the moving equipments, given that
we can know the transformation rule $T_{\mathbf{V}}$ independently
of knowing the operational definitions of $\xi'_{1},\xi'_{2},\ldots\xi'_{n}$.
Typically, $T_{\mathbf{V}}$ is thought to be derived from the assumption
that the RP \eqref{eq:RP-math1a} holds. If however $M_{\mathbf{V}}$
is, by definition, equal to $T_{\mathbf{V}}^{-1}\circ P_{\mathbf{V}}$,
then in place of \eqref{eq:RP-math1a} we have the tautology (\ref{eq:RP-tautologi}),
which does not determine $T_{\mathbf{V}}$.
\item [{(d)}] Therefore, unsurprisingly, it is not the RP from which the
transformation rules are routinely deduced, but the covariance \eqref{eq:kovi-a}.
As we have seen, however, covariance is, in general, neither sufficient
nor necessary for the RP. Whether \eqref{eq:RP-math1a} implies \eqref{eq:kovi-a}
hinges on the physical fact whether \eqref{eq:RE-kov2} is satisfied.
But, if $M_{\mathbf{V}}$ is taken to be $T_{\mathbf{V}}^{-1}\circ P_{\mathbf{V}}$
by definition, the RP becomes true---in the form of tautology (\ref{eq:RP-tautologi})---but
does not imply covariance $T_{\mathbf{V}}^{-1}\circ P_{\mathbf{V}}(\mathcal{E})=\mathcal{E}$.
\item [{(e)}] Even if we assume that a {}``transformation rule'' function
$\phi'\circ T_{\mathbf{V}}\circ\phi^{-1}$ were derived from some
independent premises---from the independent assumption of covariance,
for example---how do we know that the $T_{\mathbf{V}}$ we obtained
and the quantities of values $\phi'\circ T_{\mathbf{V}}\circ\phi^{-1}\left(\xi_{1},\xi_{2},\ldots\xi_{n}\right)$
are correct plugins for the RP? How could we verify that $\phi'\circ T_{\mathbf{V}}\circ\phi^{-1}\left(\xi_{1},\xi_{2},\ldots\xi_{n}\right)$
are indeed the values measured by a moving observer applying the same
operations with the same measuring equipments, etc.?---without having
an independent concept of $M_{\mathbf{V}}$, at least for the measuring
equipments?
\item [{(f)}] One could argue that we do not need such a verification;
$\phi'\circ T_{\mathbf{V}}\circ\phi^{-1}\left(\xi_{1},\xi_{2},\ldots\xi_{n}\right)$
can be regarded \emph{as the empirical definition} of the primed quantities:
\begin{equation}
\left(\xi'_{1},\xi'_{2},\ldots\xi'_{n}\right)\overset{^{def}}{=}\phi'\circ T_{\mathbf{V}}\circ\phi^{-1}\left(\xi_{1},\xi_{2},\ldots\xi_{n}\right)\label{eq:trafo-mint-def}
\end{equation}
This is of course logically possible. The operational definition of
the primed quantities would say: ask the observer at rest in $K$
to measure $\xi_{1},\xi_{2},\ldots\xi_{n}$ with the measuring equipments
at rest in $K$, and then perform the mathematical operation \eqref{eq:trafo-mint-def}.
In this way, however, even the transformation rules would become tautologies;
they would be true, no matter how the things are in the physical world.
\item [{(g)}] Someone might claim that the identity of $M_{\mathbf{V}}$
with $T_{\mathbf{V}}^{-1}\circ P_{\mathbf{V}}$ is not a simple stipulation
but rather an analytic truth which follows from the identity of the
two \emph{concepts}. Still, if that were the case, RP would be a statement
which is true in all possible worlds; independently of any contingent
fact of nature; independently of the actual behavior of moving physical
objects.
\item [{(h)}] On the contrary, as we have already pointed out in Remark~\ref{Meg:kontingencia},
$M_{\mathbf{V}}(F)$ and $T_{\mathbf{V}}^{-1}\left(P_{\mathbf{V}}(F)\right)$
are \emph{different concepts}, referring to different features of
different parts of the physical reality. Any connection between the
two things must be a contingent fact of the world. 
\item [{(i)}] $T_{\mathbf{V}}^{-1}\circ P_{\mathbf{V}}$ is a $2^{R}\rightarrow2^{R}$
map which is completely determined\emph{ }by\emph{ }the physical behaviors
of the\emph{ measuring} equipments. On the other hand, whether the
elements of $\mathcal{E}\subset2^{R}$ satisfy condition (M) and whether
$T_{\mathbf{V}}^{-1}\circ P_{\mathbf{V}}(\mathcal{E})\subseteq\mathcal{E}$
depend on the actual physical properties of the \emph{object} physical
system.
\item [{(j)}] Let us note that in the standard textbook applications of
the RP $M_{\mathbf{V}}$ is used as an independent concept, without
any prior reference to the Lorentz boost $T_{\mathbf{V}}^{-1}\circ P_{\mathbf{V}}$.
For example, we do not need to refer to the Lorentz transformations
in order to understand the concept of `the stationary electromagnetic
field of a uniformly moving point charge'; as we are capable to solve
the electrodynamical equations for such a situation, within one single
frame of reference, without even knowing of the Lorentz transformation
rules.\hfill{} $\lrcorner$\medskip{}

\end{lyxlist}

\section*{Acknowledgment}

The research was partly supported by the OTKA Foundation, No.~K 68043.

\section*{References}

~
\begin{lyxlist}{00.00.0000}
\item [{Bell,~J.S.~(1987):}] How to teach special relativity, in \emph{Speakable
and unspeakable in quantum mechanics}. Cambridge, Cambridge University
Press.
\item [{Einstein,~A~(1905):}] \foreignlanguage{ngerman}{Zur Elektrodynamik
bewegter Körper, \emph{Annalen der Physik}} \textbf{17}, 891. (On
the Electrodynamics of Moving Bodies, in H. A. Lorentz et al.,\emph{
The principle of relativity: a collection of original memoirs on the
special and general theory of relativity. }London, Methuen and Company
1923)
\item [{Georgiou,~A.~(1969):}] Special relativity and thermodynamics,
\emph{Proc. Comb. Phil. Soc.} \textbf{66}, 423. 
\item [{Jackson,~J.D.~(1999):}] \emph{Classical Electrodynamics (Third
edition).} Hoboken (NJ), John Wiley \& Sons.
\item [{Szabó,~L.E.~(2004):}] On the meaning of Lorentz covariance, \emph{Foundations
of Physics Letters} \textbf{17}, pp. 479--496.
\item [{Tolman,~R.C.~(1949):}] \emph{Relativity, Thermodynamics and Cosmology.}
Oxford, Clarendon Press.\end{lyxlist}

\end{document}